\documentclass[%
 reprint,
 superscriptaddress,
 amsmath,amssymb,
 prb,
]{revtex4-2}

\usepackage{graphicx}
\usepackage{dcolumn}
\usepackage{bm}
\usepackage{graphicx}
\usepackage{color}
\usepackage[colorlinks, linkcolor=blue, anchorcolor=blue, citecolor=blue, urlcolor=blue]{hyperref}
\usepackage{amsmath}
\usepackage{multirow}
\usepackage{tabularx}
\usepackage{rotating}
\emergencystretch=\maxdimen
\newcommand{\cmao}{CeMgAl$_{11}$O$_{19}$}
\newcommand{\lzao}{LaZnAl$_{11}$O$_{19}$}
\newcommand{\pmao}{PrMgAl$_{11}$O$_{19}$}
\newcommand{\nzao}{NdZnAl$_{11}$O$_{19}$}

\begin{document}


\title{Magnetic ground state and persistent spin fluctuations in triangular-lattice antiferromagnet NdZnAl$_{11}$O$_{19}$}



\author{Yantao Cao}
\affiliation{Institute of Physics, Chinese Academy of Sciences, Beijing 100190, China}
\affiliation{Songshan Lake Materials Laboratory, Dongguan 523808, China}
\affiliation{School of Physical Sciences, University of Chinese Academy of Sciences, Beijing 101408, China}

\author{Huanpeng Bu}
\affiliation{Songshan Lake Materials Laboratory, Dongguan 523808, China}

\author{Toni Shiroka}
\affiliation{PSI Center for Neutron and Muon Sciences CNM, Paul Scherrer Institut, CH-5232 Villigen PSI, Switzerland}
\affiliation{Laboratorium f\"{u}r Festk\"{o}rperphysik, ETH Z\"{u}rich, CH-8093 Zurich, Switzerland}

\author{Helen C. Walker}
\affiliation{ISIS Neutron and Muon Source, Rutherford Appleton Laboratory, Chilton, Didcot OX11 0QX, United Kingdom}

\author{Zhendong Fu}
\affiliation{Songshan Lake Materials Laboratory, Dongguan 523808, China}

\author{Zhaoming Tian}
\email{tianzhaoming@hust.edu.cn}
\affiliation{School of Physics and Wuhan National High Magnetic Field Center,
Huazhong University of Science and Technology, Wuhan 430074, P. R. China.}

\author{Jinkui Zhao}
\email{jkzhao@iphy.ac.cn}
\affiliation{Institute of Physics, Chinese Academy of Sciences, Beijing 100190, China}
\affiliation{Songshan Lake Materials Laboratory, Dongguan 523808, China}
\affiliation{School of Physical Sciences, University of Chinese Academy of Sciences, Beijing 101408, China}
\affiliation{School of Physical Sciences, Great Bay University, Dongguan 523808, China}

\author{Hanjie Guo}
\email{hjguo@sslab.org.cn}
\affiliation{Songshan Lake Materials Laboratory, Dongguan 523808, China}



\date{\today}

\begin{abstract}
Rare-earth triangular-lattice magnets serve as an excellent platform for investigating exotic quantum magnetic phenomena. Recently, the hexaaluminate \cmao\ has been proposed to host a $U(1)$ Dirac quantum spin liquid state with dominant Ising anisotropy. Here, we report a systematic study of its analogue, \nzao, employing ac susceptibility, inelastic neutron scattering, and muon spin relaxation measurements. Inelastic neutron scattering measurements establish a well-defined $J_\mathrm{eff}$ = 1/2 ground state with moderate Ising anisotropy ($g_c$ = 4.54, $g_\mathrm{ab}$ = 1.42). Muon spin relaxation measurements reveal persistent fluctuations emerging below $\sim$15\,K, and extending down to at least 0.28 K. AC susceptibility data further indicate an absence of magnetic ordering or spin freezing down to 50\,mK, despite an overall antiferromagnetic interaction with the Curie-Weiss temprature of $-0.42$\,K. These results suggest that \nzao\ is a good candidate material for realizing a quantum spin liquid state.
\end{abstract}


\maketitle

\section{Introduction}
A quantum spin liquid (QSL) is an intriguing state where electronic spins are highly correlated, but resist ordering even at zero temperature due to strong quantum fluctuations \cite{balents2010spin,savary2016quantum,broholm2020quantum}. Since Anderson's pioneering work, QSLs have attracted widespread attention \cite{anderson1987resonating,anderson1973resonating}. However, due to the influence of defects and site disorder etc., no real material has been universally accepted as a definitive realization of a QSL.

Triangular-lattice antiferromagnets have long been a fertile ground to search for exotic quantum states. A nearest-neighbor Heisenberg model predicts a 120$^\circ$ order both classically and quantum mechanically \cite{Capriotti1999}. However, interactions beyond nearest neighbors \cite{Iqbal2016} or anisotropy \cite{Wannier1950} may still give rise to a spin liquid state. Recent studies on neodymium heptatantalate NdTa$_7$O$_{19}$ \cite{Arh2022} reveal Ising-like anisotropy and a possible QSL state. However, only polycrystals or small single crystals \cite{Lia2025} were available for this system, hindering further studies using techniques such as inelastic neutron scattering.

Another promising system with Ising-like anisotropy is the layered rare-earth hexaaluminate $R$(Mg/Zn)Al$_{11}$O$_{19}$ ($R$ = rare earth) \cite{ashtar2019,Bu2022,cao2025,cao2024,Li2024,Ma2024}. For example, directional magnetization studies on \cmao\ unraveled a large anisotropy with $g_c/g_{ab} \sim$ 9.5 \cite{cao2025}. Furthermore, detailed specific heat and muon spin relaxation ($\mu$SR) measurements suggested a $U(1)$ Dirac QSL state for \cmao. One advantage of this Mg-containing system is the availability of large single crystals grown by optical floating zone technique. Nonetheless, it is more difficult to grow single crystals of the Zn-containing samples due to the relatively stronger volatility of the Zn element compared to Mg.

In this work, we report a comprehensive study of \nzao\ polycrystals using magnetic susceptibility, inelastic neutron scattering (INS) and $\mu$SR measurements. Compared to isostructural \cmao, this compound is less anisotropic, but still shows a moderate Ising-like anisotropy with $g_c/g_{ab} \sim$ 3.2. The large gap of the first crystal-electric-field (CEF) excitation ensures that the low temperature properties could be described by a $J_\mathrm{eff}$ = 1/2 state. The system shows persistent spin fluctuations below about 15\,K, but no ordering nor spin freezing down to 50\,mK, making it a plausible candidate to realize a QSL state.

\section{Experimental Methods}

Polycrystalline samples of \nzao\ were prepared using a standard solid-state reaction method. Raw materials of Nd$_2$O$_3$ (99.99\%), ZnO (99.99\%), and Al$_2$O$_3$ (99.99\%) were dried at 900$^{\circ}$C overnight to avoid moisture contamination. Then, stoichiometric amounts of them were mixed and ground thoroughly, pressed into pellets, and sintered at 1550$^{\circ}$C with several intermediate grindings. No impurity phases were detected within the accuracy of laboratory X-ray diffraction measurements \cite{supp}.

DC and AC magnetic susceptibility measurements above 2\,K were performed on a Physical Property Measurement System (PPMS, Quantum Design) equipped with a vibrating sample magnetometer (VSM) and an ACMS-II option. AC susceptibility down to 50\,mK was measured on the PPMS with a dilution insert.

Inelastic neutron scattering measurements were performed on the MERLIN spectrometer at ISIS, UK. The samples were loaded into aluminium foil sachets, which were then wrapped around the inner part of a cylindrical aluminium can and cooled down to 7\,K by a close-cycled refrigerator. MERLIN was operated in multirep mode with incident neutron energies of 23.0, 36.5, 67.1 and 160.6\,meV. Data \cite{data1,data2} were processed using MANTID \cite{mantid}. The phonon signal was removed from the Nd sample data using isostructural nonmagnetic La sample data scaled appropriately for the relative sample masses.

Zero-field (ZF) and longitudinal-field (LF) muon spin relaxation ($\mu$SR) measurements were performed on the Dolly spectrometer at the Paul Scherrer Institut (PSI), Switzerland. The asymmetry is defined as $A(t) = [F(t) - \alpha B(t)]/[F(t) + \alpha B(t)]$, where $F(t)$ and $B(t)$ are the number of positrons hitting the forward and backward detectors at time \textit{t}, respectively. The parameter $\alpha$ reflects the relative counting efficiency of the two detectors and the sample geometry. The $\mu$SR spectra were analyzed using the MUSRFIT program \cite{suter2012musrfit}.

\section{Results and Discussions}

\begin{figure}
\includegraphics[width=.45\textwidth]{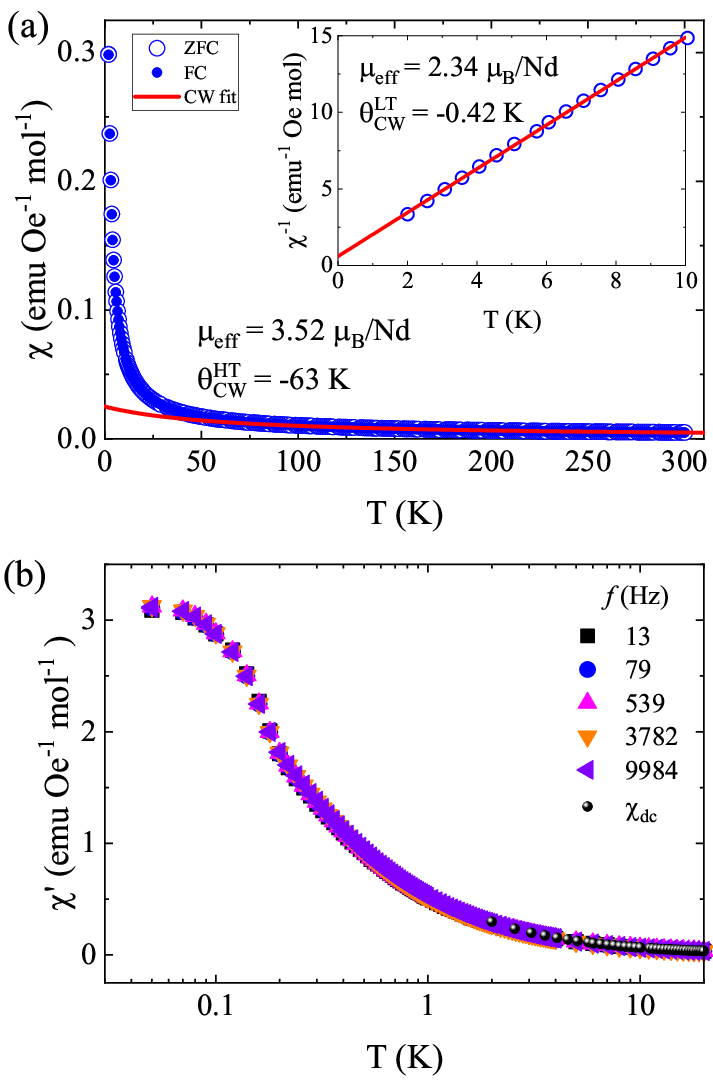}
\caption{(a) Temperature dependence of the magnetic susceptibility, $\chi$, measured in the ZFC and FC modes at a field of 0.1 T. The insets shows the inverse susceptibility at low temperatures. The red curves are CW fits. (b) Temperature dependence of the ac susceptibility measured at various frequencies. The dc susceptibility is also shown for comparison. }
\label{sus}
\end{figure}

Figure \ref{sus}(a) shows the temperature dependence of the DC magnetic susceptibility measured in the zero-field-cooled (ZFC) and field-cooled (FC) modes. No bifurcation or difference between the ZFC and FC curves is observed down to 2\,K, indicating a paramagnetic state without spin freezing above this temperature. A Curie-Weiss (CW) fit, $\chi = \chi_0 + C/(T-\theta_\mathrm{CW}^\mathrm{HT})$ to the data between 150 and 300\,K yields a negative Weiss temperature $\theta_\mathrm{CW}^\mathrm{HT}$ of $-63(1)$\,K, and an effective moment $\mu_\mathrm{eff}$ = $\sqrt{8C}$ of 3.53(2)\,$\mu_\mathrm{B}$/Nd. This moment is close to the expected value for the Nd$^{3+}$ free ion (3.62\,$\mu_\mathrm{B}$), whereas the Weiss temperature should have substantial impact from the CEF effect. To estimate the interactions among the Nd$^{3+}$ spins, the susceptibility below 10\,K was analyzed with $\chi^{-1}(T) = (T-\theta_\mathrm{CW}^\mathrm{LT})/C$. The extracted $\theta_\mathrm{CW}^\mathrm{LT}$ amounts to $-0.42(3)$\,K, indicating an overall antiferromagnetic interaction. The averaged exchange interaction could be estimated as $\bar{J} = \frac{2}{3}\theta_\mathrm{CW}^\mathrm{LT}$ = $-0.28(2)$\,K. Fig. \ref{sus}(b) depicts the temperature dependence of the real component of the ac susceptibility, $\chi'$, down to 50\,mK. $\chi'$ increases continuously with decreasing temperature and levels off below $\sim$0.1\,K. No signature of magnetic ordering is observed down to 50\,mK. Moreover, the frequency independency of $\chi'$ rules out a spin glass state.

\begin{figure}
\includegraphics[width=.48\textwidth]{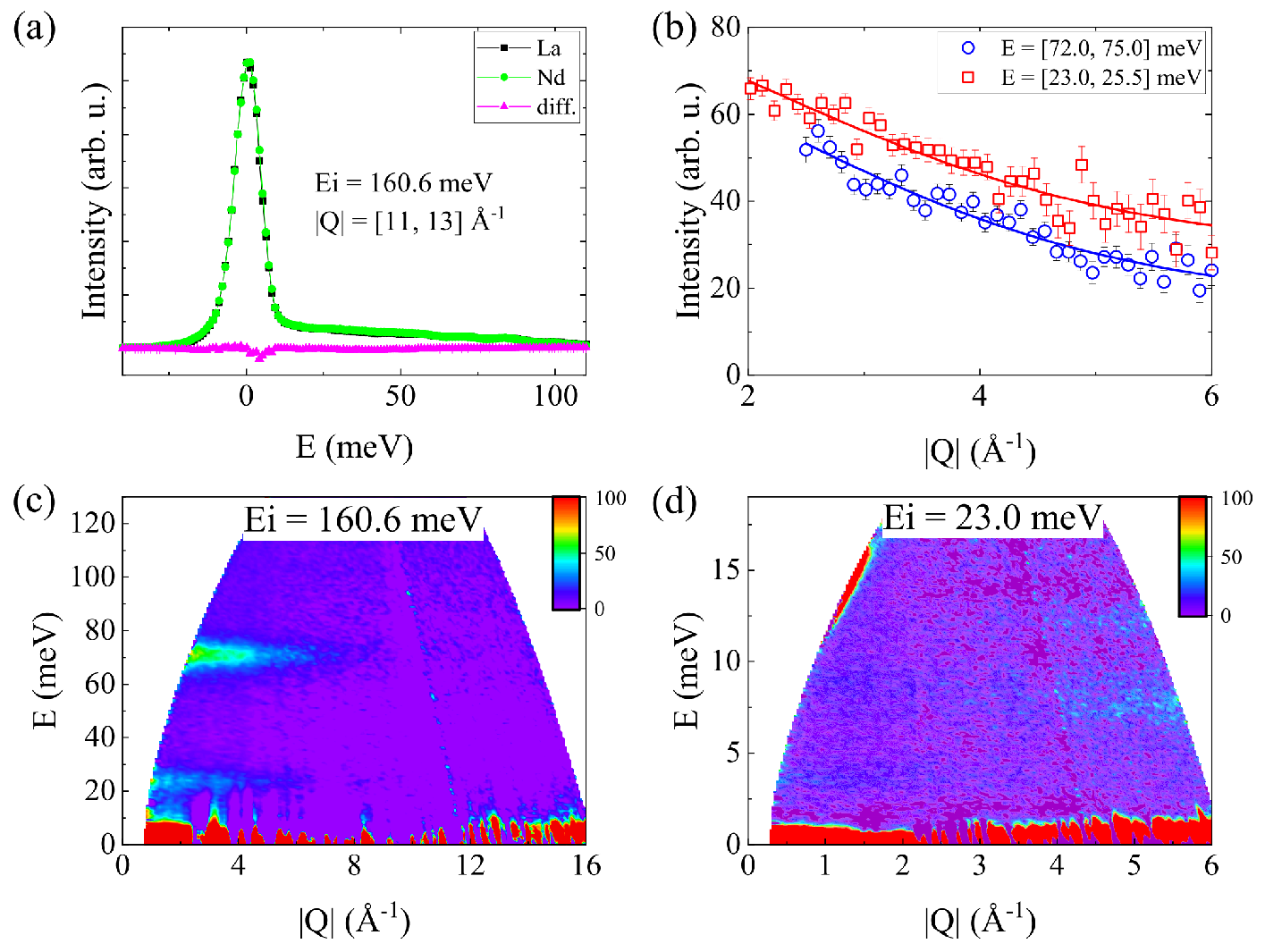}
\caption{(a) High-$Q$ cuts of the INS data for \lzao\ and \nzao. The \lzao\ data has been re-scaled to match the intensity of \nzao. The bottom curve shows the difference between the two datasets. (b) $Q$ dependence of the integrated intensities for the excitations at $E$ = 23.8 and 70.7\,meV. (c) and (d) INS spectra of \nzao\ at 7\,K with incident neutron energies $E_i$ = 160.6 and 23.0\,meV, respectively. The phonon contributions have been subtracted using the \lzao\ data.}
\label{neutron}
\end{figure}

To determine the CEF scheme of Nd$^{3+}$ in \nzao, we performed inelastic neutron scattering measurements. The phonon contributions were carefully subtracted using the scaled La sample data. The scale was determined by the high-$|Q|$ data such that the two datasets overlap with each other; see Fig. \ref{neutron}(a). Two dispersionless excitations at 23.8 and 70.7\,meV can be observed in Fig. \ref{neutron}(c). The momentum transfer ($Q$) dependence of the integrated intensities for these two excitations follows the form factor of Nd$^{3+}$; see Fig. \ref{neutron}(b). At lower energies, no obvious CEF excitations can be observed, as shown in Fig. \ref{neutron}(d).  Note that the intense localized spot at the low-$|Q|$ edge is a spurion. Some weak intensities, which increase with increasing $|Q|$, at around 7.5 and 12.5\,meV are most probably phonons due to imperfect subtractions. Two additional weak dispersionless excitations are discernible at $\sim$3.5 and 9\,meV. These two excitations have much broader peak width compared to the instrumental resolution, we thus attribute them to the small amount of Nd$^{3+}$ ions displaced to the 6$h$ site \cite{supp}.

\begin{figure*}
\includegraphics[width=.98\textwidth]{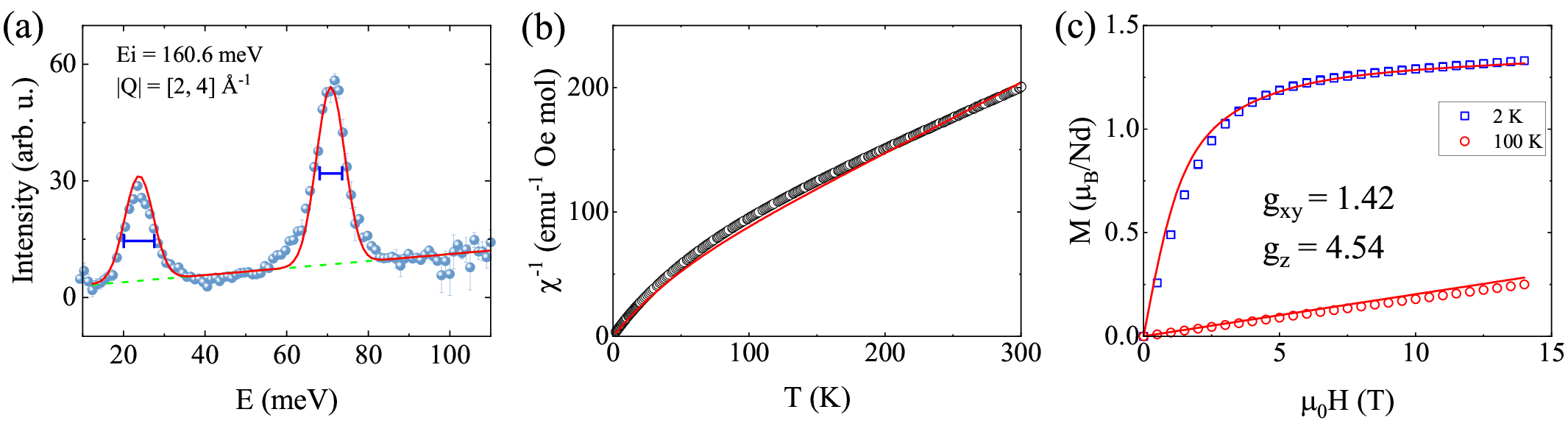}
\caption{CEF fits to (a) the INS data at 7\,K and (c) the magnetization data at 2\,K. The green dashed line in (a) is a linear background. The blue horizontal bars are the instrument resolution at the corresponding energies. The calculated inverse susceptibility is shown in (b) along with the experimental data. }
\label{CEF}
\end{figure*}

The majority of the Nd$^{3+}$ ions are located at the 2\textit{d} site with a point group $D_{3h}$. The crystal field splits the Hund's rule ground-state multiplet $^{4}I_{9/2}$ into five Kramers doublet in the paramagnetic state. Thus, four transitions from the ground state to the excited states are expected at low temperatures. However, only two peaks are observed experimentally with their peak widths comparable with the instrumental resolution; see Fig. \ref{CEF}(a). The other two should either have very weak intensities or reside at higher energies. The absence of observable peaks could introduce ambiguity in determining the CEF scheme. To analyze the data, we start with a weak coupling scheme acting on the $|J,m_J\rangle$ basis using the PyCrystalField software \cite{Scheie2021}. The CEF Hamiltonian can be expressed as
\begin{equation}\label{eq1}
  \mathcal{H}_\mathrm{CEF} = \sum B_l^mO_l^m,
\end{equation}
where $O_l^m$ are the Stevens operators and $B_l^m$ are the CEF parameters. With the quantization axes along the $\bar{6}$ axis, only the $B_2^0$,  $B_4^0$, $B_6^0$, and $B_6^6$ parameters are nonzero. The presence of the $B_6^6$ term mixes those states differing in $m_J$ by six. Thus, the CEF eigenfunctions are of the form such as $|\pm1/2\rangle$ and sin$\alpha$$|\mp9/2\rangle$+cos$\alpha$$|\pm3/2\rangle$ etc. These states have distinct anisotropies. Therefore, a simultaneous fit to the INS and magnetization data may overcome the obstacle to determining the CEF scheme. To calculate the magnetization, a Zeeman term, $g_J\mu_B\mu_0\textbf{H}\cdot \textbf{J}$, was added to Eq. \ref{eq1} with \textbf{H} along different directions. The powder averaged magnetization was then calculated as $M_\mathrm{CEF} = 2/3M_{x} + 1/3M_z$. In addition, specific heat data indicate an excitation at $\sim$9.5\,meV \cite{supp}, which provides another constraint on the eigenvalues.

The powder averaged neutron cross section with dipole approximation can be expressed as \cite{Boothroyd}
\begin{equation}\label{}
\begin{split}
  \frac{d\sigma^2}{d\Omega d\omega} = & N(\gamma r_0)^2\frac{k_f}{k_i}F^2(Q)\mathrm{exp}[-2W(Q)]\sum_{i,j} p_i\times \\ &\frac{2}{3}\sum_\alpha |\langle \Gamma_j|\hat{J}_\alpha|\Gamma_i\rangle|^2\delta(\hbar\omega + E_i - E_j),
\end{split}
\end{equation}
where $\alpha$ = \textit{x}, \textit{y} and \textit{z}, and the other symbols have generic meanings. The best fits to the INS data at 7\,K and isothermal magnetization data at 2\,K are shown in Fig. \ref{CEF}(a) and (c). Based on the refined model, the inverse susceptibility and isothermal magnetization at 100\,K were calculated and shown to agree well with experimental values; see Fig. \ref{CEF}(b) and (c). The small discrepancy in the susceptibility around 100 K could be due to the omission of the contributions from the Nd$^{3+}$ ions at the $6h$ site. The corresponding eigenvalues and eigenvectors are summarized in Tab. \ref{EV1}. The overall energy scale within the $^4I_{9/2}$ multiplet is similar to that of many other Nd systems \cite{Xu2015,Anand2017,Scheie2018}. The ground state doublet is dominated by the $|m_J = \pm7/2\rangle$ state with a small contribution from the $|m_J = \mp5/2\rangle$ state. The \textit{g} factors corresponding to the ground state are $g_{xy}$ = 1.42 and $g_z$ = 4.54. Also note that the transition from the ground state ($GS$) to the first and third excited states ($ES_{1,3}$) has small intensities because the matrix element of $\langle GS|\hat{J}_\alpha|ES_{1,3}\rangle$ is either zero or a small number. This explains why only two excitations are observed from neutron scattering.

\begin{table*}
\caption{CEF eigenvectors and eigenvalues for \nzao\ in the weak coupling scheme ($|Jm_J\rangle$ basis). The CEF parameters are $B20$ = -0.20(4), $B40$ = -0.0033(2), $B60$ = 0.00056(2) and $B66$ = 0.0034(4)\,meV.}
\begin{ruledtabular}
\begin{tabular}{c|cccccccccc}
E (meV) &$| -\frac{9}{2}\rangle$ & $| -\frac{7}{2}\rangle$ & $| -\frac{5}{2}\rangle$ & $| -\frac{3}{2}\rangle$ & $| -\frac{1}{2}\rangle$ & $| \frac{1}{2}\rangle$ & $| \frac{3}{2}\rangle$ & $| \frac{5}{2}\rangle$ & $| \frac{7}{2}\rangle$ & $| \frac{9}{2}\rangle$ \tabularnewline
 \hline
0.000 & 0.0 & 0.9678 & 0.0 & 0.0 & 0.0 & 0.0 & 0.0 & -0.2517 & 0.0 & 0.0 \tabularnewline
0.000 & 0.0 & 0.0 & 0.2517 & 0.0 & 0.0 & 0.0 & 0.0 & 0.0 & -0.9678 & 0.0 \tabularnewline
9.400 & 0.0 & 0.0 & 0.0 & 0.0 & 1.0 & 0.0 & 0.0 & 0.0 & 0.0 & 0.0 \tabularnewline
9.400 & 0.0 & 0.0 & 0.0 & 0.0 & 0.0 & 1.0 & 0.0 & 0.0 & 0.0 & 0.0 \tabularnewline
23.788 & -0.9282 & 0.0 & 0.0 & 0.0 & 0.0 & 0.0 & 0.3722 & 0.0 & 0.0 & 0.0 \tabularnewline
23.788 & 0.0 & 0.0 & 0.0 & -0.3722 & 0.0 & 0.0 & 0.0 & 0.0 & 0.0 & 0.9282 \tabularnewline
56.456 & 0.0 & 0.0 & 0.0 & 0.9282 & 0.0 & 0.0 & 0.0 & 0.0 & 0.0 & 0.3722 \tabularnewline
56.456 & 0.3722 & 0.0 & 0.0 & 0.0 & 0.0 & 0.0 & 0.9282 & 0.0 & 0.0 & 0.0 \tabularnewline
70.760 & 0.0 & -0.2517 & 0.0 & 0.0 & 0.0 & 0.0 & 0.0 & -0.9678 & 0.0 & 0.0 \tabularnewline
70.760 & 0.0 & 0.0 & -0.9678 & 0.0 & 0.0 & 0.0 & 0.0 & 0.0 & -0.2517 & 0.0 \tabularnewline \tabularnewline
\end{tabular}\end{ruledtabular}
\label{EV1}
\end{table*}

The Ising anisotropy of \nzao\ is weaker than that of CeMgAl$_{11}$O$_{19}$ \cite{cao2025} or PrMgAl$_{11}$O$_{19}$ \cite{cao2024} due to a mixing of the $|m_J = \pm7/2\rangle$ and $|m_J = \mp5/2\rangle$ states in the ground state. The large first excitation gap ($\sim$110\,K) ensures that the low temperature properties are governed by the ground state doublet.

\begin{figure}
\includegraphics[width=.45\textwidth]{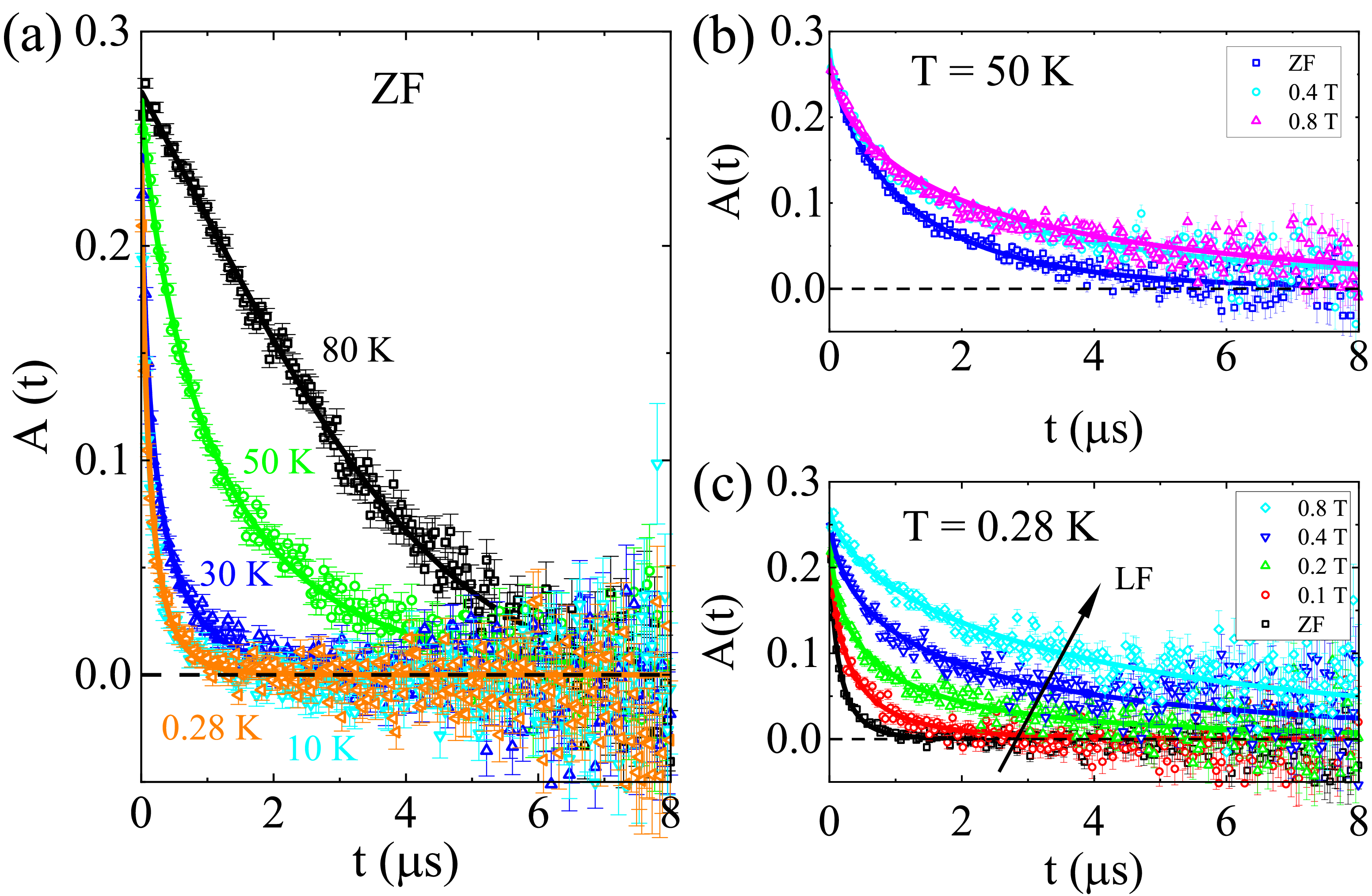}
\caption{(a) Typical ZF-$\mu$SR spectra measured at various temperatures. LF dependence of the $\mu$SR spectra measured at 50 and 0.28\,K are shown in (b) and (c), respectively.}
\label{musr}
\end{figure}

To probe the spin dynamics at low temperatures, we performed $\mu$SR measurements down to 0.28\,K.
Fig. \ref{musr}(a) presents some typical ZF-$\mu$SR spectra measured between 0.28 and 80\,K. No spontaneous muon spin precession is observed, ruling out any long-range magnetic ordering down to 0.28\,K. Fig. \ref{musr}(b) shows the LF dependence of the spectra at 50\,K. The spectra at 0.4 and 0.8\,T are almost identical, indicating that the electron spins fluctuate at a rate much larger than $\gamma_\mu B_\mathrm{LF}$, where $\gamma_\mu/2\pi$ = 135.5\,MHz/T is the gyromagnetic ratio of muon. The asymmetry is slightly recovered compared to that at ZF, indicating that there exists a small static component, most probably from the Al and Nd nuclear spins. At 0.28\,K, however, the spectrum exhibits clear dependence on LF, suggesting that the fluctuation of electron spins has slowed down. All the spectra can be well described by the stretched exponential function:

\begin{equation}
A(t)=A_1 \exp[-(\lambda t)^{\beta}].
\label{muSr}
\end{equation}
The temperature dependence of the relaxation rate $\lambda$ and exponent $\beta$ at ZF is presented in Fig. \ref{parameter}(a). With decreasing temperatures, $\lambda$ increases steeply and tends to saturate below $\sim$15\,K. Such behavior can be phenomenologically described by $\lambda^{-1} = \lambda_0^{-1} + C \mathrm{exp}(-\delta/k_B T)$, as shown in Fig. \ref{parameter}(a). The best fit yields $\delta$ = 14.7(3)\,meV and $\lambda_0$ = 8.5(1)\,$\mu s^{-1}$. The value of $\delta$ agrees reasonably well with the first CEF excitation level extracted from the specific heat data \cite{supp}. In the fast fluctuation regime, the muon spin relaxation rate is proportional to the inverse of the fluctuation rate of electron spin, i.e., $\lambda \propto 1/\nu$. Therefore, the thermally activated behavior can be ascribed to the Orbach process via the first CEF level \cite{orbach1961} in this temperature range, as observed in other rare earth systems \cite{Lago2007,Khasanov2008}, reflecting the thermally activated behavior of the fluctuation rate $\nu \propto \mathrm{exp}(-\delta/k_B T)$. However, it should be noted that the implanted muons could potentially have a slight influence on the CEF levels \cite{wu2025}.
The continuous increase of $\lambda$ with decreasing temperature indicates a slowing down of the spin fluctuations. At lower temperatures, the system enters a quantum regime where relaxation occurs within the ground state doublet. The constant behavior of $\lambda$ at lower temperatures is consistent with that observed in many frustrated systems \cite{Uemura1994,Khuntia2016,Yang2024}, suggesting the persistence of fluctuations even close to the absolute zero. Note that the plateau onset temperature from $\mu$SR is much larger than that from the ac susceptibility measurement. This could be partially understood by the different time scales probed by $\mu$SR and ac susceptibility techniques, which are roughly sensitive to dynamics in the $\mu$s and ms time scale, respectively. On the other hand, it could also be due to the different portion of the $Q$ space probed by the two techniques. As a bulk probe, ac susceptibility is sensitive to dynamics at $Q$ = 0, whereas $\mu$SR as a local probe is sensitive to dynamics summed over all $Q$s.

The exponent $\beta$ also shows a temperature dependent behavior at high temperatures, evolving from $\sim$1 at 80\,K to $\sim$0.6 at 15\,K. It then also levels off at lower temperatures; see Fig. \ref{musr}(a). A $\beta$ value less than 1 would indicate a distribution of the relaxation rates that could be attributed to the anisotropic magnetism and averaged over all directions in a polycrystal, or due to multiple muon stopping sites which become inequivalent at low temperatures as the spin fluctuations slow down.

\begin{figure}
\includegraphics[width=.45\textwidth]{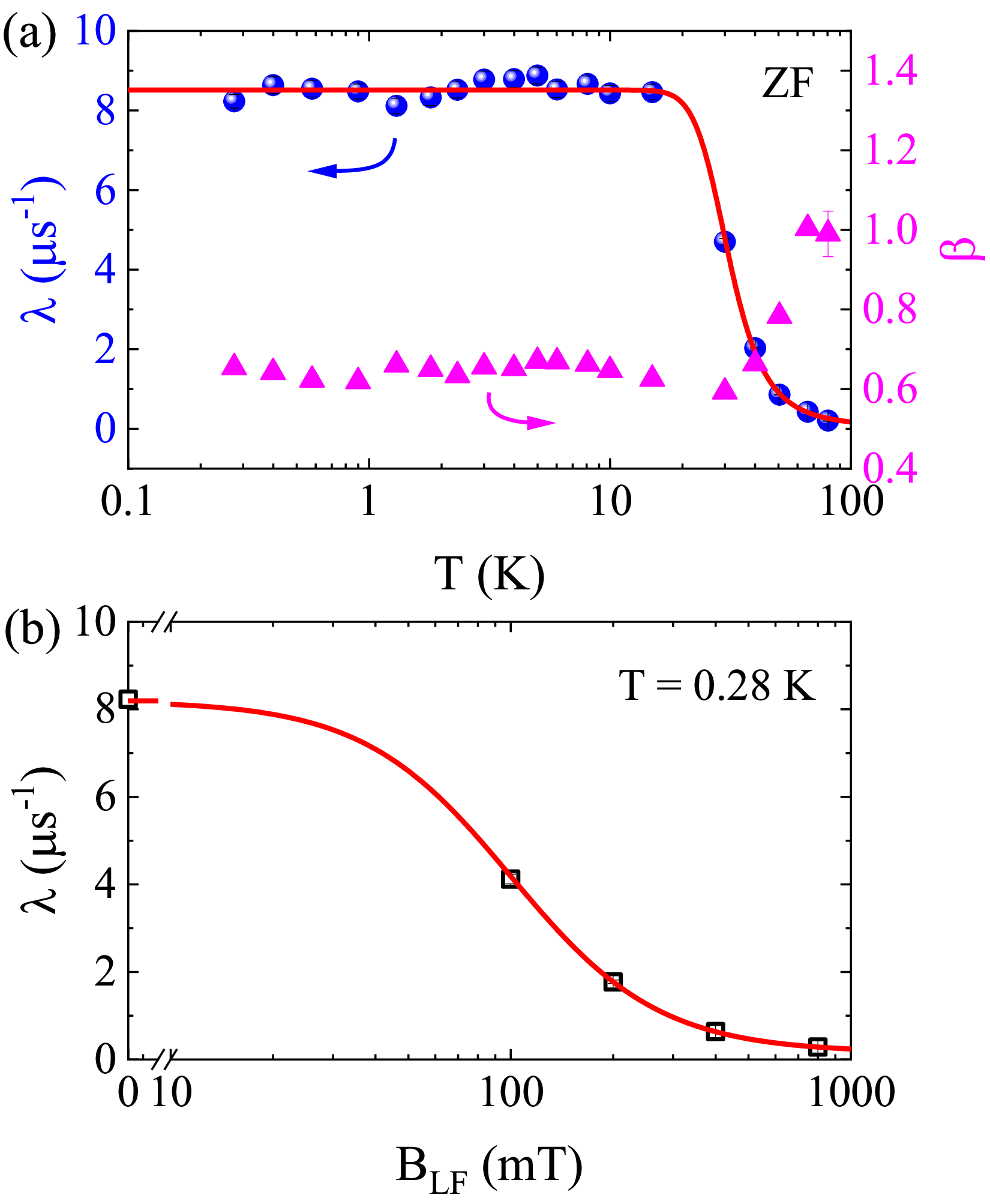}
\caption{ (a) Temperature dependence of the extracted muon spin relaxation rate $\lambda$ and exponent $\beta$. The red curve is a fit as described in the text. (b) Longitudinal field dependence of the relaxation rate at 0.28\,K. The red curve is a fit according to Eq.  \ref{Redfield}. }
\label{parameter}
\end{figure}

The spin fluctuation rate was further probed by the LF measurements. In the motional narrowing limit ($\nu \gg \gamma_\mu \Delta$), the LF dependence of the relaxation rate can be described by the modified Redfield formula \cite{guo2014}:

\begin{equation}
\lambda(B_\mathrm{LF})=\frac{2(\gamma_\mu\Delta)^2\nu}{\nu^2+(\gamma_{\mu} B_\mathrm{LF})^2} + \lambda_0,
\label{Redfield}
\end{equation}
where $\Delta$ is the rms of the internal field distribution width. The deduced parameters are $\Delta$ = 21.76(8)\,mT, $\nu$ = 85.3(8)\,MHz, and $\lambda_0$ = 0.160(4)\,$\mu$s$^{-1}$, respectively. The ratio of $\nu$/($\gamma_\mu \Delta$) amounts to 4.5, consistent with the motional narrowing assumptiom. The fluctuation rate is about 2 orders larger than that of \cmao \cite{cao2025}.  This could partly be due to a higher temperature probed here. However, the temperature independency of $\lambda$ below 15\,K suggests that such fast fluctuations could persist to even lower temperatures. Note that \nzao\ and \cmao\ exhibit distinct anisotropic behavior. The ratio of $g_c/g_{ab}$ estimated from CEF analysis amounts to 9.5 for \cmao, while it is only 3.2 for \nzao. The more XY-like character for \nzao\ may facilitate quantum fluctuations. Although the fluctuation rate is large compared to that of \cmao, it is still much smaller than the frequency of exchange fluctuations, which is estimated to be $|k_\mathrm{B}\bar{J}/\hbar|$ = 37\,GHz, reflecting a highly correlated character of the spins.

Interestingly, the magnetic properties of \nzao\ are reminiscent of that for NdTa$_7$O$_{19}$ \cite{Arh2022} in several aspects, reflected in the anisotropy of the $g$ factors, the onset temperature of the plateau in relaxation rate, and the Curie-Weiss temperature at low temperatures. These similarities indicate that \nzao\ could be another promising system to realize a QSL state dominated by Ising anisotropy. Nonetheless, there are some differences that warrant further elaboration.
The CEF ground state of \nzao\ is dominated by the $|m_J = \pm7/2\rangle$  state, while $|m_J = \pm5/2\rangle$ dominates in the NdTa$_7$O$_{19}$ case. The larger $m_J$ value results in a larger moment, and thus a larger relaxation rate (Fig. \ref{parameter}(a)). Another important issue is that
in this rare-earth hexaaluminate system, a certain degree of disorder at the 4\textit{e} and 2\textit{d} sites has been discovered \cite{cao2024,cao2025}. Disorder may strongly affect the magnetic properties of non-Kramers ions such as Pr$^{3+}$, but for Kramers ions such as Nd$^{3+}$, the magnetic ground state is more robust due to the protection of time reversal symmetry. As shown in Fig. \ref{neutron}(a), the CEF peak widths are comparable to the instrumental resolution, suggesting that disorder may only play a minor role. This is also corroborated by the ac susceptibility measurements, which show the absence of spin glass transition down to 50 mK. Considering the overall antiferromagnetic coupling with $\theta_\mathrm{CW}$ = $-0.42$\,K, the frustration index is more than $|-0.42/0.05|$ = 8.4. Along with the persistent fluctuations as evidenced by $\mu$SR measurement, all these results suggest that \nzao\ may host a quantum spin liquid state. Future single crystal synthesis and neutron scattering measurements to quantify the exchange energy and search for excitation continuum are highly desirable.

\section{CONCLUSIONS}

We performed a detailed analysis of the magnetic ground state of \nzao. INS measurements reveal a ground state doublet well separated from the first excited state, validating a $J_\mathrm{eff}$ = 1/2 picture at low temperatures. $\mu$SR measurements unravel persistent fluctuations of the Nd$^{3+}$ moments down to at least 0.28 K. AC susceptibility data additionally rule out any spin glass transition in this system. All these results indicate that this system may be another good candidate to realize a quantum spin liquid state with dominant Ising anisotropy.

\begin{acknowledgments}

We thank Wei Li and Yuan Gao for valuable discussions. We also thank Duc Le for suggestions on estimating the uncertainty of the fitting parameters. This work was supported by the Guangdong Basic and Applied Basic Research Foundation (Grant No. 2022B1515120020). We gratefully acknowledge the Science and Technology Facilities Council (STFC) for Xpress access to neutron beam time on MERLIN at ISIS. Part of this work was based on experiments performed at the Swiss Muon Source S$\mu$S, Paul Scherrer Institute, Villigen, Switzerland.

\end{acknowledgments}

\bibliography{ref}

\setcounter{figure}{0}
\setcounter{table}{0}
\section*{Supplementary Materials}
\renewcommand{\thefigure}{S\arabic{figure}}
\renewcommand{\thetable}{S\arabic{table}}

\begin{figure}
\includegraphics[width=.45\textwidth]{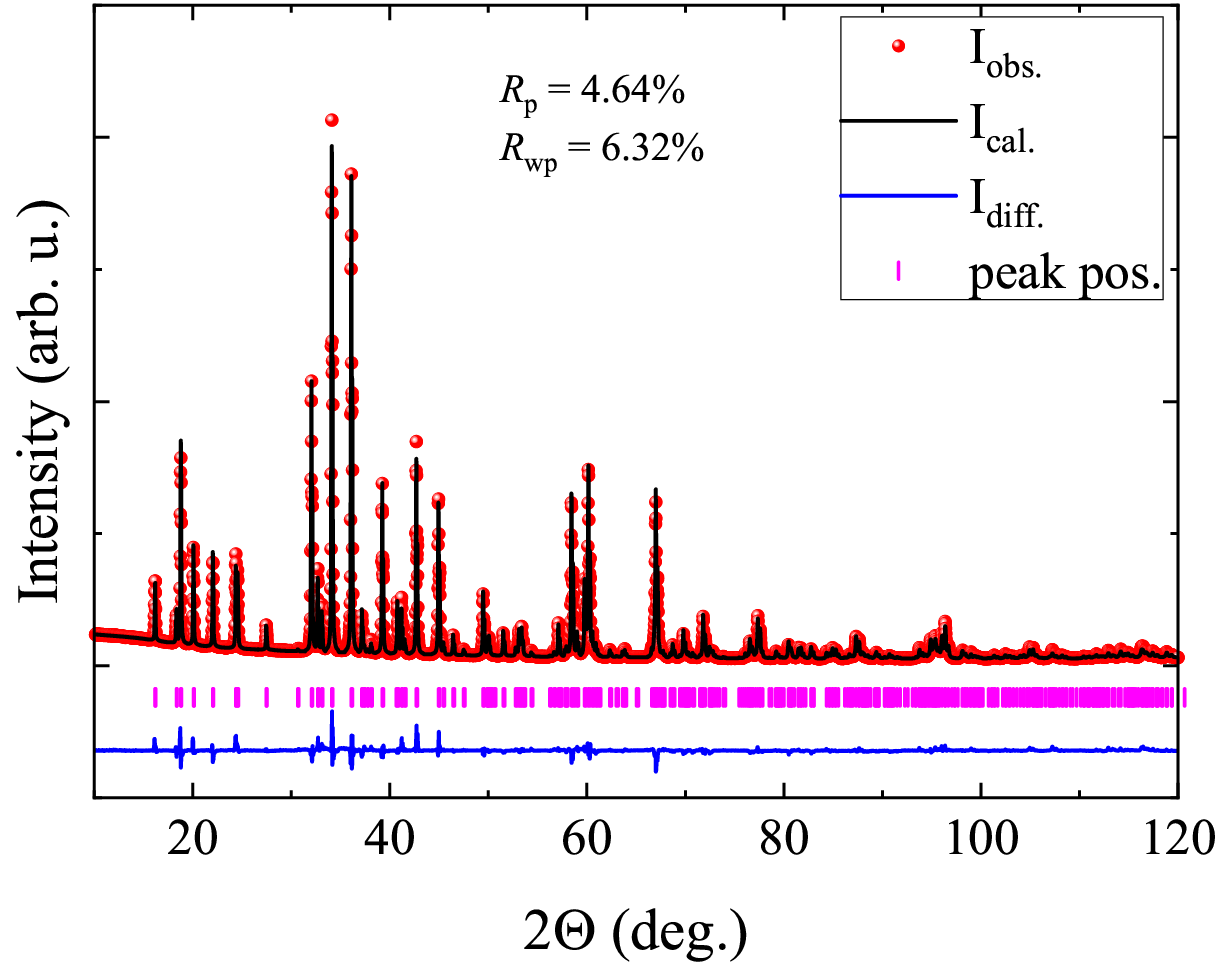}
\caption{Rietveld refinement against the powder XRD pattern for \nzao\ collected at room temperature.}
\label{xrd}
\end{figure}

X-ray diffraction (XRD) measurements were performed on a Rigaku Miniflex with Cu $K_\alpha$ radiation at room temperature. The refined pattern is shown in Fig. \ref{xrd}. The refined crystal structural parameters are listed in Tab. \ref{parameter}. The structure is consistent with that observed in \pmao\ \cite{cao2024} and \cmao\ \cite{cao2025} with certain degree of disorder at the $2d$ and $6h$ sites. Nevertheless, a single crystal study is required to better quantify the degree of disorder in \nzao.

\begin{figure}
\includegraphics[width=.45\textwidth]{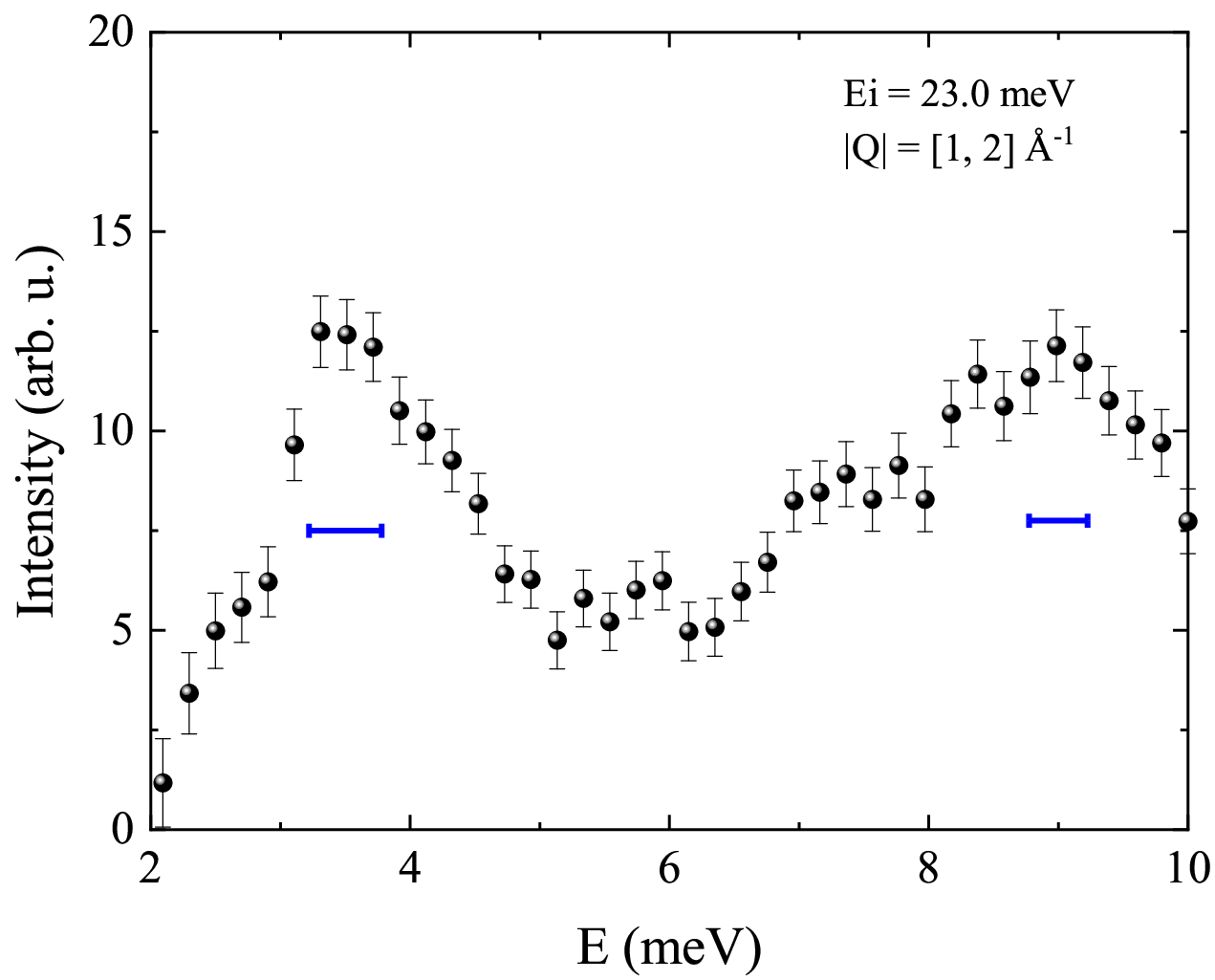}
\caption{Energy dependence of the integrated intensities for \nzao\ measured at 7 K. The blue bars represent the instrumental resolution (full width at half maximum).}
\label{ins}
\end{figure}
\begin{table*}
\caption{Refined structural parameters for \nzao\ at room temperature. The space group is $P6_3/mmc$ (No. 194). The lattice constants are $a = b =$ 5.58432(4) \AA, and $c$ = 21.91109(18) \AA. \label{parameter}}
\begin{tabularx}{\linewidth}{X X X X X X}
\hline
\hline
Atom  & occ. &  $x$ & $y$ & $z$ & $B$(\AA)  \\
Nd1 (2$d$)  &  0.780(5)   & 1/3 & 2/3  & 0.75 & 0.906(38)  \\
Nd2 (6$h$)  &  0.072(2) &  0.25201(104) & 0.74799(104) & 0.75 & 0.906(38) \\
Al1 (12$k$)  & 1   &  0.16841(39) & 0.33677(78) & 0.60877(8) & 1.945(40)          \\
Al2 (4$f$)  &  1   & 1/3 & 2/3 & 0.19081(14) & 2.188(77)   \\
Al3 (4$f$)  & 0.5 &  1/3 & 2/3 & 0.47229(10) & 2.589(50)  \\
Mg  (4$f$)  & 0.5 &  1/3 & 2/3 & 0.47229(10) & 2.589(50)  \\
Al4 (2$a$) &  1   & 0 & 0 &0 & 2.494(138)  \\
Al5 (4$e$)  & 0.5 &  0   &  0  &  0.24501( 113) & 2.788(202) \\
O1  (6$h$)  & 1   &  0.18063(105) & 0.36123(210) & 1/4  & 2.005(133) \\
O2  (4$f$) & 1   & 0.33333 & 0.66667  &0.56079(26) & 2.505(157) \\
O3  (12$k$) & 1   & -0.00406(67) & 0.49797(33) & 0.34888(16) & 1.956(88) \\
O4  (12$k$)  & 1   &  0.15009(70) & 0.30011(140) & 0.44672(13)&  1.782(86)  \\
O5  (4$e$)  & 1   &  0      & 0 &  0.35049(36)  & 3.025(187)  \\
\hline
\hline
\end{tabularx}
\end{table*}

\begin{figure}
\includegraphics[width=.45\textwidth]{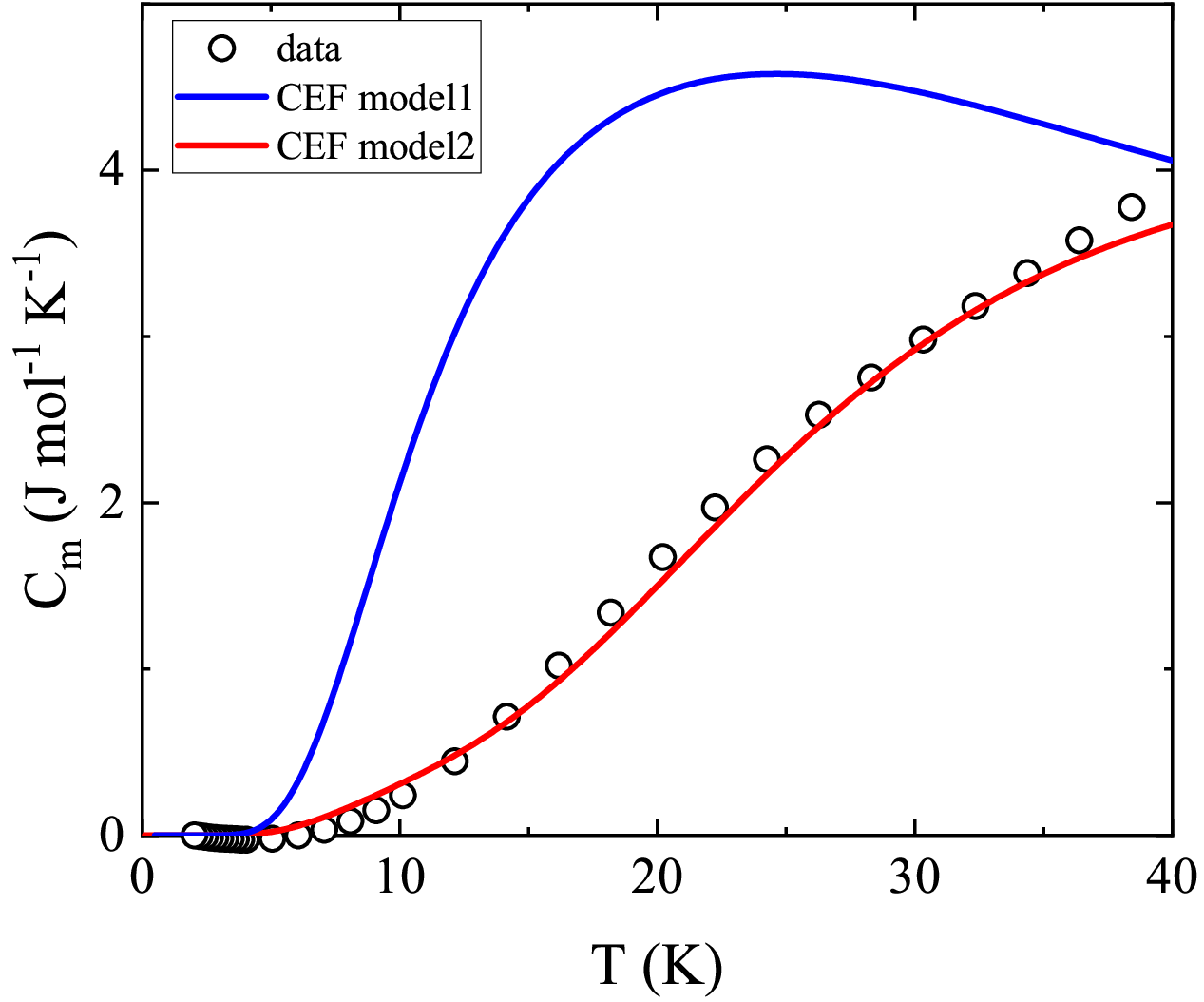}
\caption{Temperature dependence of the magnetic specific heat for \nzao\ and comparison with CEF model calculations.}
\label{HC}
\end{figure}

Figure \ref{ins} presents the energy dependence of the integrated INS intensities with incident neutron energy of 23.0\,meV. In contrast to the excitations at 23.8 and 70.7\,meV, these two peaks are much weaker and broader than the instrumental resolution. We thus ascribe them to the CEF excitations related to the minor Nd$^{3+}$ ions at the $6h$ site. This assignment is also corroborated by the specific data. Fig. \ref{HC} shows the temperature dependence of the magnetic specific heat for \nzao\ obtained by subtracting the phonon contributions using the La-sample data. The CEF contributions to the specific heat can be calculated as \cite{Gopal}:
\begin{equation}\label{schottky}
  C_\mathrm{CEF} = R\beta^2[\frac{\sum_i E_i^2 n_i e^{-\beta E_i}}{Z} - (\frac{\sum_i E_i n_i e^{-\beta E_i}}{Z})^2],
\end{equation}
where $R$ is the ideal gas constant, $\beta = 1/k_B T$, $n_i$ is the degeneracy of energy level $E_i$, and $Z$ is the partition function.
Suppose the excitations at 3.5, 9.0, 23.8 and 70.7\,meV are all from Nd$^{3+}$ ions at the $2d$ site, the calculated specific heat is shown in Fig. \ref{HC} as a blue curve (denoted as model1). Obviously, this model overestimates the specific heat below 40 K, indicating that the 3.5-meV and 9.0-meV excitations should have less weights. The structure refinement indicated two distinct Nd sites. Thus, it is plausible that these two weak excitations originate from the minor Nd ions at the $6h$ sites. Therefore, a weighted model (denoted as model2) is constructed as
\begin{equation}\label{schottky}
  C_\mathrm{total} = fC_\mathrm{CEF}^{2d} + (1-f)C_\mathrm{CEF}^{6h}.
\end{equation}
where $C_\mathrm{CEF}^{6h}$ includes excitations at 3.5 and 9.0\,meV, while in $C_\mathrm{CEF}^{2d}$ the high energy excitation energies are fixed, but the first excitation energy, $E_1$ is set as a free parameter. The best fit yields $f$ = 0.86(1), and $E_1$ for the $2d$ site is 9.5(1)\,meV, which is used to fit the INS spectrum in the main text. The obtained $f$ agrees reasonably well with the occupancy of the $2d$ site from the structure refinement.

\end{document}